\documentclass[showpacs,amsmath,amssymb,twocolumn,prl,superscriptaddress, nobibnotes]{revtex4-1}
\usepackage{amssymb}
\usepackage{graphicx}
\usepackage{enumerate}
\usepackage{epsfig}
\usepackage{subfigure}
\usepackage{xcolor}
\usepackage[T1]{fontenc}
\usepackage{textcomp}
\usepackage{fullpage}
\usepackage{amsthm,amsfonts,amssymb,amscd,mathrsfs,xspace,framed}
\usepackage{amsmath}
\usepackage{color}
\usepackage{setspace}
\usepackage{url}
\usepackage{wrapfig}
\usepackage{enumitem}
\begin{document}

\title{Noise suppression of quantum frequency conversion in thin-film lithium niobate waveguides via adapted poling}

\author{Ian Briggs}
\affiliation{Chandra Department of Electrical and Computer Engineering, The University of Texas at Austin}
\affiliation{Wyant College of Optical Sciences, The University of Arizona}
\author{Pao-Kang Chen}
\affiliation{Chandra Department of Electrical and Computer Engineering, The University of Texas at Austin}
\author{Jiaqi Huang}
\affiliation{Chandra Department of Electrical and Computer Engineering, The University of Texas at Austin}
\author{Linran Fan}
\email{linran.fan@utexas.edu}
\affiliation{Chandra Department of Electrical and Computer Engineering, The University of Texas at Austin}

\maketitle

\textbf{
Quantum frequency conversion (QFC) enables optical interfaces between dissimilar quantum systems, but suppressing noise while maintaining high conversion efficiency remains challenging. Here, we demonstrate that adapted poling in thin-film lithium niobate (TFLN) waveguides reduces the pump-power requirement while shifting the dominant noise mechanism and its power scaling. In 2-cm-long waveguides, adapted poling enables maximum conversion efficiency at an on-chip pump power of $19\pm0.8$~mW, reducing the required pump power by 8~dB and the total conversion noise by 14~dB relative to conventional periodic poling at their respective maximum-conversion operating points. Controlled measurements with different pump conditions and device configurations reveal a transition from predominantly quadratic noise scaling in periodically poled devices to predominantly linear scaling in adapted-poling devices. This transition accompanies a shift from noise dominated by cascaded second-harmonic generation and Stokes spontaneous Raman scattering to direct anti-Stokes spontaneous Raman scattering. These results establish adapted poling as an effective approach to suppressing nonlinear noise, enabling low-noise integrated frequency converters for quantum networks.
}

Hybrid quantum systems combine the complementary capabilities of distinct physical platforms, offering a promising route to scalable quantum networks~\cite{Wallquist:2009,doi:10.1073/pnas.1419326112}. Matter-based systems, such as neutral atoms, trapped ions, color centers, and quantum dots, are indispensable for storing and processing quantum information~\cite{Specht:11,Blatt:08,Abobeih_2018,Gao:2015}. However, their optical transitions are constrained by their energy-level structures, leading to wavelength mismatches both across platforms and with the telecom bands in fiber networks. 
Bridging these mismatches requires efficient and low-noise quantum frequency conversion (QFC)~\cite{kumar1990quantum}. Periodically poled lithium niobate (PPLN) waveguides have become a workhorse for this task, providing strong second-order nonlinearity and flexible phase matching over a broad wavelength range~\cite{Pelc:11,Maring:2018,PhysRevApplied.9.064031,Wang:2023}. 
The operating point for maximum conversion can be readily achieved by increasing the pump power.
However, suppressing the accompanying noise remains challenging. At maximum conversion, the noise is governed largely by the material’s intrinsic properties and the pump power required to reach this operating point.

TFLN waveguides have emerged as a promising platform for nonlinear optical applications, with tight optical confinement enabling strong nonlinear interactions~\cite{Zhu:21}. However, this enhanced nonlinearity does not automatically translate into low-noise operation in QFC. Thickness variations in the lithium niobate film can disrupt the phase matching along long waveguides, limiting coherent conversion and increasing the pump power required to reach maximum efficiency~\cite{adapted, Kao:26}. The stronger pump, in turn, amplifies parasitic noise processes. Maintaining phase matching over extended interaction lengths is therefore essential to reducing both the pump-power requirement and the accompanying noise in TFLN QFC waveguides.

\begin{figure*}
\centering
\includegraphics[width=\textwidth]{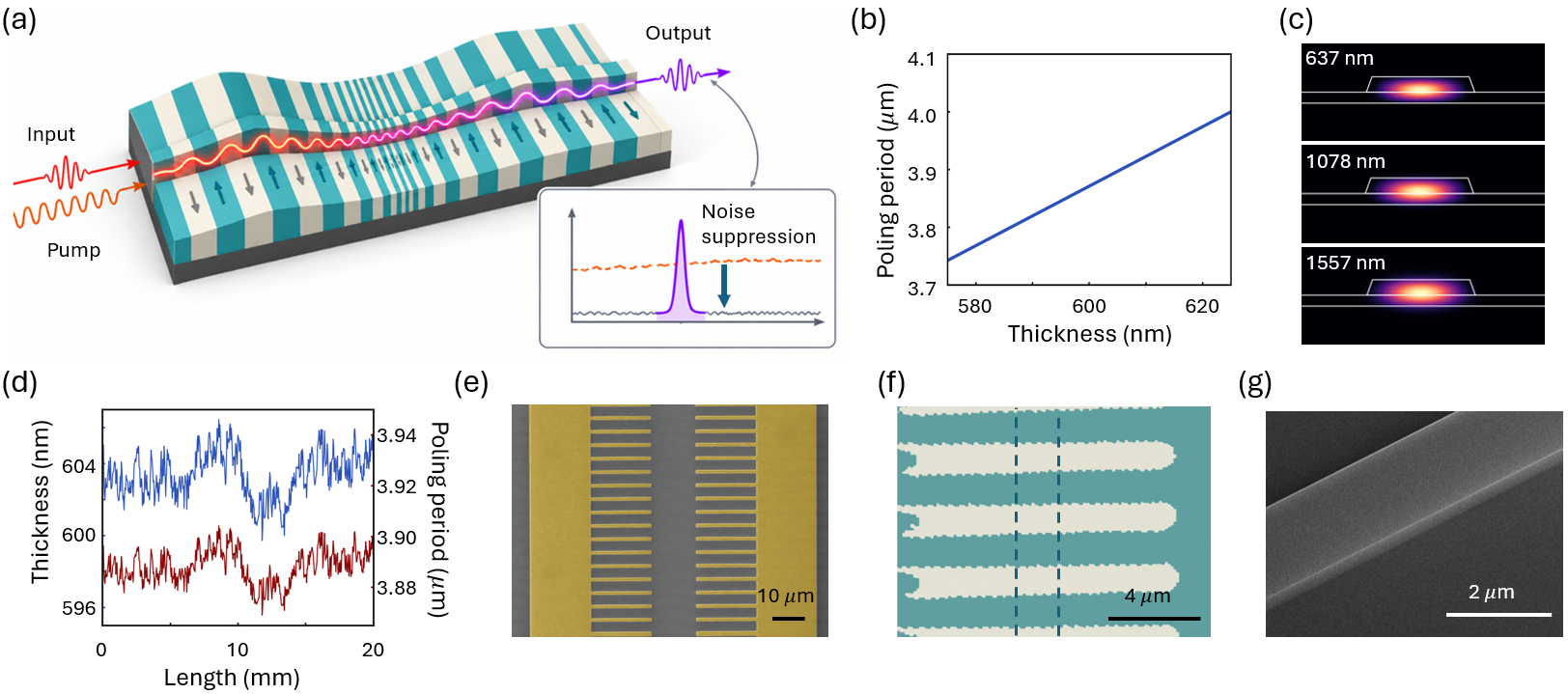}
\caption{\textbf{TFLN QFC waveguides with adapted poling.} 
(a) Illustration of TFLN QFC waveguide with adapted poling. Inset: QFC noise suppression via adapted poling. 
(b) Simulated poling period dependence on TFLN thickness for QFC between 637~nm and 1078~nm. 
(c) Simulated optical mode profiles at pump 1557~nm, input signal 1078~nm, and converted signal 637~nm. 
(d) Measured longitudinal thickness profile (blue) and corresponding local poling period (orange).
(e) False-color scanning electron micrograph of electrodes (orange) for domain inversion; scale bar $10~\mu\mathrm{m}$.
(f) Piezoresponse force microscopy (PFM) image of domain inversion in TFLN; scale bar $4~\mu\mathrm{m}$. 
(g) Scanning electron micrograph of a fabricated waveguide; scale bar $2~\mu\mathrm{m}$.}
\label{Fig1}
\end{figure*}

In this work, we demonstrate the reduction of pump power and associated noise in TFLN QFC waveguides using adapted poling.
Beyond reducing the required pump power, adapted poling shifts the dominant noise contribution from cascaded nonlinear processes to direct anti-Stokes Raman scattering, accompanied by a change from predominantly quadratic to predominantly linear noise scaling with pump power.
We target conversion between 637~nm and 1078~nm, corresponding to the zero-phonon lines of nitrogen-vacancy centers in diamond and neutral divacancy centers in 4H silicon carbide, respectively, using a pump near 1557~nm (Fig.~\ref{Fig1}(a))~\cite{janitz2020cavity, castelletto2020silicon}. These defect systems are selected for their promising performance as optical quantum memories. 
The devices are fabricated on a nominal 600-nm-thick X-cut lithium-niobate-on-insulator wafer. 
To use the largest $d_{33}$ nonlinear coefficient, waveguides are aligned along the Y direction of lithium niobate, and the fundamental transverse-electric (TE) modes are used for all wavelengths (Fig.~\ref{Fig1}(c)). Waveguides have top width of $1.55~\mu\mathrm{m}$, etch depth of 350~nm, and a sidewall angle of $55^\circ$. 
With 600 nm thickness, the poling period is simulated as 3.87~$\mu$m. The poling period shifts over 50~nm with a thickness change of 10~nm (Fig.~\ref{Fig1}(b)).


Before fabrication, we measure the lithium niobate film thickness along the propagation direction using a white-light reflectometer mounted on a motorized stage. 
With this representative device, the film thickness varies between 600~nm and 606~nm across the 2-cm waveguide length (Fig.~\ref{Fig1}(d)). 
The local poling period is adjusted between 3.87~$\mu$m and 3.9~$\mu$m accordingly. The domain inversion of lithium niobate is achieved by applying high-voltage electric pulses through on-chip electrodes (Fig.~\ref{Fig1}(e)).
We verify the domain inversion using piezoresponse force microscopy (Fig.~\ref{Fig1}(f)) and subsequently fabricate  waveguides by argon-based inductively coupled plasma reactive-ion etching (Fig.~\ref{Fig1}(g)). Besides waveguides with adapted poling, we also fabricate devices with periodic poling on the same chip for comparison.

\begin{figure*}[t]
\centering
\includegraphics[width=\textwidth]{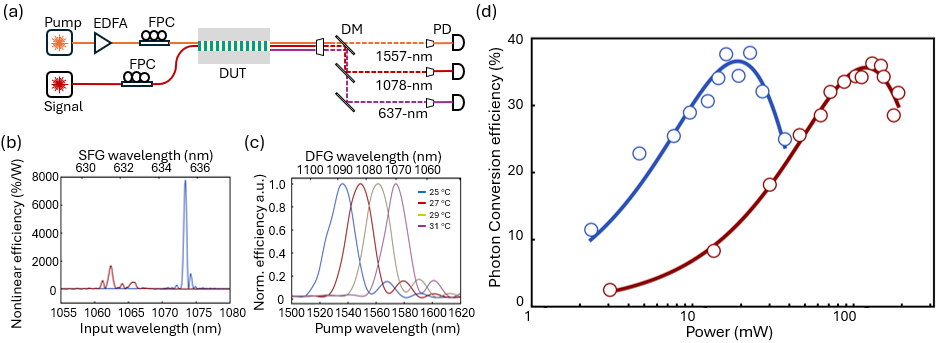}
\caption{\textbf{Conversion efficiency measurements.} 
(a) Experimental setup. EDFA: erbium-doped fiber amplifier; FPC: fiber polarization controller; DUT: device under test; DM: dichroic mirror; PD: photodetector. Solid line: fiber paths; dashed lines: free-space optical paths. Orange: pump wavelength at 1557~nm; red: input signal wavelength at 1078~nm; purple: converted signal wavelength at 637~nm.
(b) Sum-frequency generation (SFG) spectra of 20-mm-long adapted-poling (blue) and periodic-poling (red) devices, measured by sweeping the signal wavelength with the pump fixed at 1557~nm.
(c) Difference-frequency generation (DFG) spectra of the adapted-poling device at temperatures from 25 to $31~^\circ\mathrm{C}$, measured by sweeping the pump wavelength with the input signal fixed at 637~nm. 
(d) On-chip conversion efficiency as a function of on-chip pump power for adapted-poling (blue) and periodic-poling (red) devices. Solid lines are fitted from measured results (circles) with the equation $\eta = \eta_\mathrm{max}\cdot\sin^2(\sqrt{\eta_\mathrm{norm}\cdot P}L)$.
}
\label{Fig2}
\end{figure*}

We first verify the frequency conversion efficiency through the sum-frequency generation (SFG) process.
The measurement setup is shown in Fig.~\ref{Fig2}(a).
With the pump wavelength fixed at 1557~nm, we scan the input signal wavelength and measure the SFG power. 
The SFG spectrum with periodic poling deviates substantially from the theoretical sinc-squared function (Fig.~\ref{Fig2}(b)). The SFG spectrum shows multiple peaks as different waveguide sections have different phase-matching conditions, with the peak nonlinear efficiency measured as $1600\pm100\%$/W. 
With the adapted poling approach, the SFG spectrum shows only one dominant peak, matching the theoretical sinc-squared function. The peak nonlinear efficiency up to $7800\pm600\%$/W is measured.
Because of propagation losses, the effective interaction length ($L_\mathrm{eff}$) is smaller than the actual device length ($L$). In the low-power regime, the ratio is
\begin{equation}
    \frac{L_{\mathrm{eff}}}{L} = \frac{1-e^{-\Gamma L}}{\Gamma L} \cdot e^{-\frac{\alpha_cL}{2}}
\end{equation}
with $\Gamma = (\alpha_p+\alpha_s-\alpha_c)/2$, and $\alpha_p$, $\alpha_s$, $\alpha_c$ the pump, input signal, and converted signal propagation losses respectively~(Supplementary Section I).
Thus, the impact of propagation losses must be considered when analyzing the nonlinear efficiency.
At 1557, 1078, and 637~nm, the measured propagation losses are $\alpha_p=0.5~\mathrm{dB}/\mathrm{cm}$, $\alpha_s=0.6~\mathrm{dB}/\mathrm{cm}$, and $\alpha_c=2.4~\mathrm{dB}/\mathrm{cm}$, respectively (See Methods).
Then the nonlinear efficiency corrected by propagation losses with the factor of $(L_\mathrm{eff}/L)^2$ reaches $17,000\pm2000\%$/W, close to the theoretical limit 18000~\%/W. This confirms that the adapted poling can recover the QFC phase matching condition disrupted by the inhomogeneous thickness of lithium niobate.
We further demonstrate that the phase matching wavelength can be adjusted by the temperature. Difference-frequency generation (DFG) spectrum is measured by fixing the input signal wavelength at 637~nm and scanning the pump wavelength.
Tuning sensitivity of $6.32\pm0.03$~nm per degree at the pump wavelength is demonstrated (Fig.~\ref{Fig2}(c)). 

Next, we examine the conversion efficiency dependence on the pump power.
For the waveguide with adapted poling, the maximum conversion is realized with an on-chip pump power of only $19\pm1$~mW (Fig.~\ref{Fig2}(d)). This value is close to the estimation of $16\pm2$~mW based on the theoretical nonlinear efficiency and measured propagation losses (Supplementary Section II).
In contrast, the maximum conversion in the waveguide with periodic poling must require pump power of $120\pm4$~mW.
Both waveguides achieve photon number conversion efficiency of $37\pm3$\%, limited by propagation losses.



\begin{figure}[t]
\centering
\includegraphics[width=0.45\textwidth]{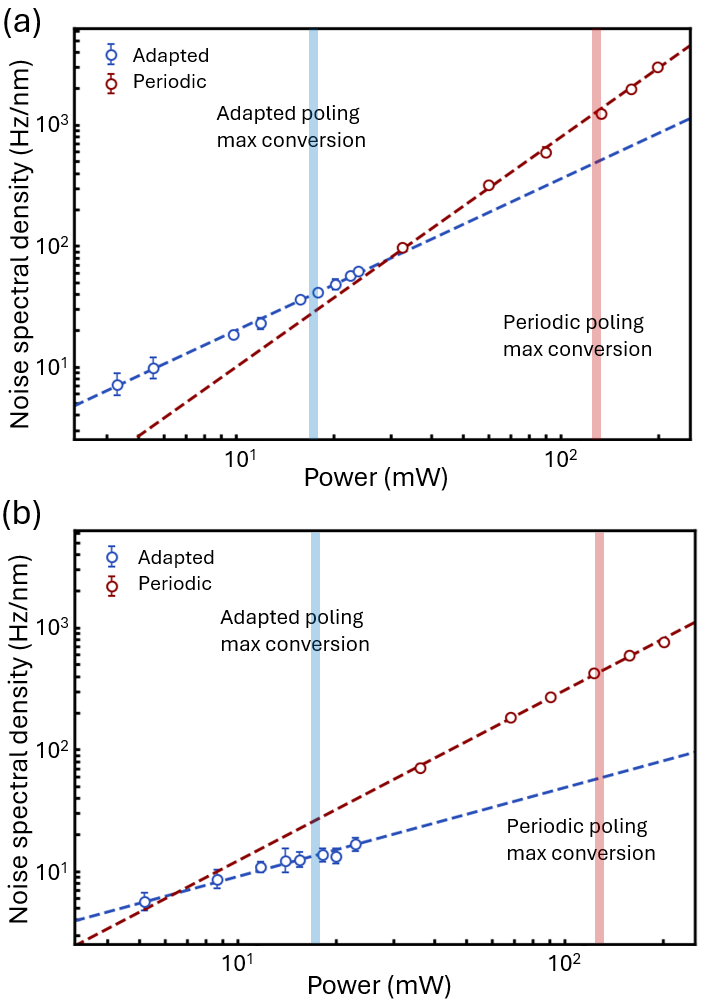}
\caption{\textbf{Total QFC noise.}
(a) Measured total noise spectral density at 1078-nm for the adapted-poling (blue) and periodic-poling (red) devices.
(b) Measured total noise spectral density at 637-nm for the adapted-poling (blue) and periodic-poling (red) devices.
Circles are measured results, and dashed lines are the linear-fit in the log-log scale. Vertical solid blue and red lines indicates the maximum conversion condition for adapted-poling and periodic-poling devices respectively.
}
\label{Fig3}
\end{figure}

The QFC noise is characterized by removing the input signal and injecting only the pump at the phase-matching wavelength, with the filtered output detected using a superconducting nanowire single-photon detector (SNSPD). Near 1078~nm, the waveguide with adapted poling exhibits an on-chip noise spectral density of $45\pm1$~Hz/nm at the maximum conversion pump power. In contrast, the noise spectral density in the waveguide with periodic poling exceeds $1.1\pm0.1$~kHz/nm, more than 14~dB higher. Log-log linear fits of noise density versus pump power yield slopes of $1.23\pm0.03$ and $1.88\pm0.05$ for adapted and periodic poling, respectively. 
This indicates that first-order noise processes dominate in the adapted-poling waveguide, whereas cascaded nonlinear processes predominate in the periodic-poling waveguide. Near 637~nm, noise spectral densities of $15\pm0.4$~Hz/nm and $400\pm15$~Hz/nm are measured in the adapted- and periodic-poling devices near maximum conversion, respectively. 

\begin{figure*}[t]
\centering
\includegraphics[width=\textwidth]{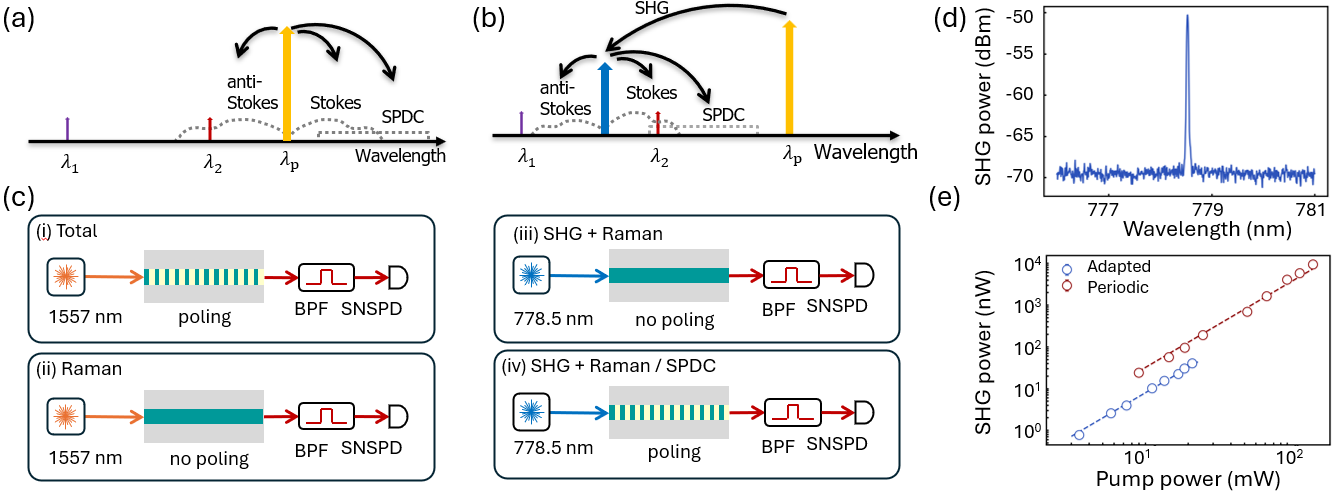}
\caption{\textbf{Methods for QFC noise analysis.}
(a) Schematic of direct noise processes, including Stokes and anti-Stokes Raman, and spontaneous parametric down-conversion (SPDC) driven by the pump.
(b) Schematic of cascaded noise processes, including Stokes and anti-Stokes Raman, and SPDC driven by the second-harmonic generation (SHG) light produced by the pump.
(c) Pump and device configurations for noise analysis. (i) pump incident on a poled QFC waveguide for total noise, (ii) pump incident on an unpoled waveguide for direct Raman noise, (iii) externally generated SHG light incident on an unpoled waveguide for cascaded SHG-Raman noise, and (iv) externally generated SHG light incident on a poled QFC waveguide for cascaded SHG-Raman and SHG-SPDC noise.
BPF: band-pass filter; SNSPD: superconducting nanowire single-photon detector.
(d) Optical spectrum of spurious SHG light. 
(e) Spurious SHG power versus pump power for the adapted-poling (blue) and periodic-poling (red) devices. Circles are measured results. Dashed lines are linear-fit in the log-log scale.
}
\label{Fig4}
\end{figure*}

We further designed different experiments to study the different noise performance in adapted-poling and periodic-poling devices.
Because the pump wavelength is longer than both signal wavelengths, neither Stokes spontaneous Raman scattering (SRS) nor direct spontaneous parametric down-conversion (SPDC) can populate the signal bands (Fig.~\ref{Fig4}(a)). Therefore, the dominant direct noise process is anti-Stokes SRS~\cite{Pelc:11,Kuo:18}.
Cascaded nonlinear processes provide additional noise pathways (Fig.~\ref{Fig4}(b))~\cite{Fan:21,Wang:2023}. Despite a large phase mismatch, the pump produces residual light near 778~nm through second-harmonic generation (SHG) (Figs.~\ref{Fig4}(d)). This SHG light subsequently generates noise in the signal bands through SRS and SPDC.

We use four pump and device configurations to isolate different noise contributions (Fig.~\ref{Fig4}(c)). In the first configuration, we measure the total noise by injecting the 1557~nm pump into a TFLN QFC device and recording the filtered output with a SNSPD. In the second configuration, we replace the QFC device with an unpoled TFLN waveguide of the same cross-section and length while retaining the pump-only input. The absence of domain inversion eliminates poling-period and duty-cycle errors, thereby suppressing the associated SHG. We therefore use this configuration to estimate the direct anti-Stokes SRS contribution.
In the third configuration, we inject SHG light generated in a commercial PPLN crystal into an unpoled TFLN waveguide. The resulting Raman noise allows us to estimate the cascaded SHG–SRS contribution. At the same SHG output power, externally injected SHG light produces a Raman noise rate three times that produced by internally generated SHG light, owing to their different longitudinal power distributions~\cite{Fan:21}. We apply this correction to infer the cascaded SHG–SRS noise during QFC.
In the fourth configuration, we inject externally generated SHG light into the TFLN QFC device, where both SRS and SPDC contribute to the measured noise. We isolate the SHG-SPDC contribution by subtracting the SHG-SRS contribution measured in the third configuration under matched SHG power. At the same SHG output power, externally injected SHG light produces an SPDC noise rate twice that produced by internally generated SH light~\cite{Wang:2023}. We apply this correction to estimate the cascaded SHG–SPDC contribution during QFC.

To relate the SHG-injection measurements to the QFC operating conditions, we independently measure the SHG output power of QFC devices as a function of the 1557~nm pump power. The SHG output exhibits the expected quadratic dependence on pump power (Fig.~\ref{Fig4}(e)). The SHG efficiencies are $8\times10^{-3}\%/\mathrm{W}$ and $3\times10^{-2}\%/\mathrm{W}$ for adapted- and periodic-poling respectively.
Matching the SHG output powers in the two measurements maps the externally injected SHG power to an equivalent fundamental pump power during QFC.

\begin{figure*}[t]
\centering
\begingroup
\setlength{\unitlength}{\dimexpr\textwidth/1000\relax}
\begin{picture}(1000,508.439)
\put(0,0){\includegraphics[width=\textwidth]{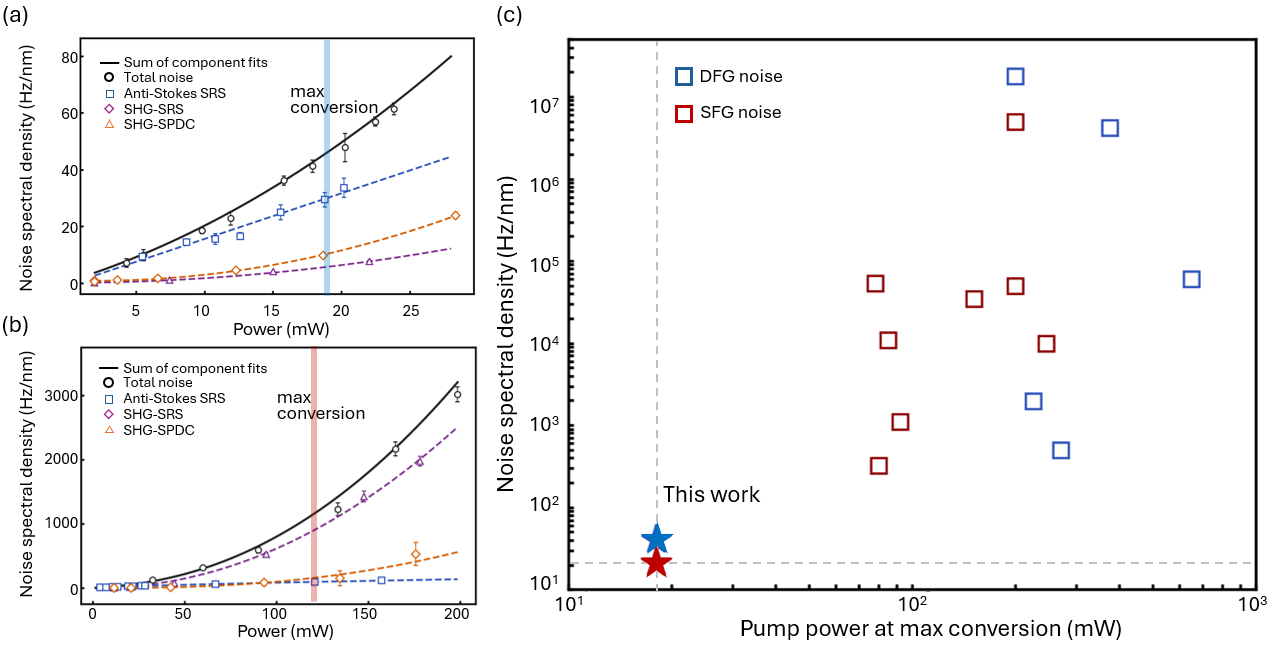}}
\put(813.50,452.85){\makebox(0,0)[l]{\normalfont\sffamily\fontsize{6.5}{7}\selectfont[\citenum{Fan:21}]}}
\put(867.90,403.23){\makebox(0,0)[r]{\normalfont\sffamily\fontsize{6.5}{7}\selectfont[\citenum{PhysRevLett.124.010510}]}}
\put(952.30,293.05){\makebox(0,0)[l]{\normalfont\sffamily\fontsize{6.5}{7}\selectfont[\citenum{photonics9120971}]}}
\put(827.80,197.05){\makebox(0,0)[l]{\normalfont\sffamily\fontsize{6.5}{7}\selectfont[\citenum{vonChamier:25}]}}
\put(849.40,158.26){\makebox(0,0)[l]{\normalfont\sffamily\fontsize{6.5}{7}\selectfont[\citenum{Yu:20}]}}
\put(813.50,416.95){\makebox(0,0)[l]{\normalfont\sffamily\fontsize{6.5}{7}\selectfont[\citenum{Fan:21}]}}
\put(703.00,289.59){\makebox(0,0)[l]{\normalfont\sffamily\fontsize{6.5}{7}\selectfont[\citenum{MA201269}]}}
\put(813.50,287.76){\makebox(0,0)[l]{\normalfont\sffamily\fontsize{6.5}{7}\selectfont[\citenum{Kuo:18}]}}
\put(755.80,268.49){\makebox(0,0)[r]{\normalfont\sffamily\fontsize{6.5}{7}\selectfont[\citenum{Pelc:11}]}}
\put(713.30,245.05){\makebox(0,0)[l]{\normalfont\sffamily\fontsize{6.5}{7}\selectfont[\citenum{Rakher:10}]}}
\put(837.80,242.41){\makebox(0,0)[l]{\normalfont\sffamily\fontsize{6.5}{7}\selectfont[\citenum{Wang:2023}]}}
\put(723.80,180.68){\makebox(0,0)[l]{\normalfont\sffamily\fontsize{6.5}{7}\selectfont[\citenum{PhysRevApplied.14.034035}]}}
\put(705.90,146.16){\makebox(0,0)[l]{\normalfont\sffamily\fontsize{6.5}{7}\selectfont[\citenum{Xu:07}]}}
\end{picture}
\endgroup
\caption{\textbf{QFC noise analysis and comparison}
(a) Noise spectral density at 1078 nm of the adapted-poling device. Black symbols: measured total noise; blue symbols: measured direct anti-Stokes Raman noise; purple symbols: measured cascaded SHG-Raman noise; orange symbols: measured cascaded SHG-SPDC noise. Blue dashed line: linear fit for direct anti-Stokes Raman noise; purple dashed line: quadratic fit for cascaded SHG-Raman noise; orange dashed line: cascaded SHG-SPDC noise. Black solid line: summation of fitted direct anti-Stokes Raman, cascaded SHG-Raman noise, and cascaded SHG-SPDC noise. The blue vertical line indicates the power for maximum conversion.
(b) Noise spectral density at 1078 nm of the periodic-poling device. Black symbols: measured total noise; purple symbols: measured cascaded SHG-Raman noise. Purple dashed line: quadratic fit for cascaded SHG-Raman noise. The red vertical line indicates the power for maximum conversion.
(c) Comparison of noise spectral density and pump power under maximum conversion for lithium niobate frequency converters with pump wavelengths longer than both signal wavelengths. 
Blue and red symbols denote the difference-frequency generation (DFG) output with longer signal wavelengths and sum-frequency generation (SFG) output with shorter wavelengths, respectively. Reference indices are labeled next to symbols.
}
\label{Fig5}
\end{figure*}

Figure~\ref{Fig5}(a) shows the noise spectral density at 1078~nm measured under the different configurations for an adapted-poling QFC waveguide. 
The direct anti-Stokes SRS from the second configuration shows linear dependence on the pump power (blue in Fig.~\ref{Fig5}(a)). 
The noise in the third configuration yields a quadratic dependence, consistent with the cascaded SHG–SRS mechanism (purple in Fig.~\ref{Fig5}(a)). 
Combining the third and fourth configurations isolates the SHG–SPDC contribution, which also scales quadratically with the pump power (orange in Fig.~\ref{Fig5}(a)). 
The sum of these independently fitted contributions closely reproduces the measured total noise rate (solid black in Fig.~\ref{Fig5}(a)) (Supplementary Section IV).
At the pump power corresponding to maximum conversion, $P=19~\mathrm{mW}$, the direct anti-Stokes SRS, SHG–SRS, and SHG–SPDC contributions are approximately $65\pm1$\%, $22\pm0.5$\%, and $13\pm0.6$\% of the measured total noise, respectively.

The same noise analysis is performed for the QFC device with periodic poling as well.  In contrast to adapted poling, cascaded processes—including SHG–SRS and SHG–SPDC—dominate the overall noise, contributing over $90\%$ of the total (Fig.~\ref{Fig5}(b)). On the other hand, the direct SRS process show noise spectral density more than one order of magnitude lower than the cascaded process. This is in agreement with the quadratic power dependence of the total noise spectral density on the pump power (Fig.~\ref{Fig5}(b)).


In the 637~nm band, both the direct anti-Stokes SRS contribution inferred from the second configuration and the cascaded SHG–SRS contribution remain below the SNSPD dark-count floor for both periodic and adapted poling devices up to the power for maximum conversion. The cascaded SHG-SPDC process does not directly contribute to the noise near 637~nm, which is shorter than the SHG wavelength at 778.5 nm.
On the other hand, only the total noise measurement in the first configuration yields a signal above this floor, indicating that upconversion of noise from the 1078~nm band dominates the 637-nm band noise. 
This interpretation agrees with the observation that the noise spectral density near 637~nm equals that near 1078 nm scaled by the conversion efficiency (Fig.~\ref{Fig3}).


We further compare the power and noise performance of the QFC waveguides with adapted poling with prior QFC works using PPLN waveguides~\cite{Fan:21,PhysRevLett.124.010510,photonics9120971,vonChamier:25,Yu:20,MA201269,Kuo:18,Pelc:11,Rakher:10,Wang:2023,PhysRevApplied.14.034035,Xu:07} (Fig.~\ref{Fig5}(c)). We only consider works where the pump wavelength is longer than both signal wavelengths to avoid unfair comparison with direct SPDC and Stokes SRS noise. We can clearly observe significant reduction of pump power over 6~dB and suppression of noise spectral density over 10~dB using the TFLN QFC devices via adapted poling.


We have demonstrated efficient frequency conversion in 20-mm-long TFLN waveguides using adapted poling to compensate for longitudinal thickness variations. 
This approach preserves coherent conversion over 2-cm-long waveguides and reduces the pump power required for maximum conversion from 120~mW to 19~mW relative to conventional periodic poling. At the respective conversion maxima, the noise spectral density decreases from 1.1~kHz/nm to 45~Hz/nm. Control measurements reveal that the lower pump-power requirement also changes the dominant noise mechanism. Direct anti-Stokes spontaneous Raman scattering dominates in the adapted-poling device near maximum conversion, producing an approximately linear dependence of noise on pump power. In contrast, the higher pump power required by the periodic-poling device drives cascaded SHG-SRS process, resulting in quadratic noise growth. Thus, the benefit of adapted poling extends beyond reducing the operating power to shifting the dominant noise mechanism and its power scaling from a predominantly quadratic to a predominantly linear regime.
These results establish the connection between thickness-compensated phase matching, reduced pump-power requirements, and suppression of nonlinear noise in integrated quantum frequency converters

\textbf{Acknowledgments}

This work was supported in part by the Defense Advanced Research Projects Agency (DARPA) HARQ program HR0011-26-9-E107, Office of Naval Research (N00014-25-1-2130), Air Force Office of Scientific Research (FA9550-24-1-0119), U.S. Department of Energy (Field Work Proposal ERKJ432). L.F. acknowledges the support from Coherent/II-VI Foundation and Sloan Fellowship. 

\vspace{4pt}
\textbf{Author contributions}

The experiments were conceived by I.B. and L.F. The device was designed and fabricated by I.B. with assistance from P.K.C. and J.H. Measurements were performed by I.B. Data analysis was conducted by I.B. and L.F. All authors participated in the manuscript preparation. L.F supervised the work.

\textbf{Methods}

\textbf{Device fabrication}
We first fabricate alignment markers for the thickness measurement, the patterning of electrodes and the patterning of photonic waveguides. After the thickness measurement, nickel electrodes are patterned on top of the lithium niobate device layer. The electrode period is obtained using data in Fig.~\ref{Fig1}(b). We apply high-voltage electric pulses to nickel electrodes for domain inversion. Then nickel electrodes are removed by hydrochloric acid. Electron-beam lithography is used to define photonic waveguides with ma-N resist. Waveguide patterns are aligned in the poled regime. Then argon-based plasma is used to transfer waveguide patterns into the lithium niobate device layer (600 nm total thickness) with 350 nm etching depth.

\textbf{Measurement setup}
The measurement setup is shown in Fig.~\ref{Fig2}(a). One tunable laser near 1078 nm (Toptica CTL-1050) is used as the signal input, and another tunable laser near 1550 nm (Santec 710) amplified by an erbium-doped fiber amplifier (EDFA) serves as the pump. The polarization is controlled by in-line fiber polarization controllers.
A fiber wavelength-division multiplexer (WDM, OZ optics) combines the pump and signal input.  A pair of lensed fibers mounted on piezoelectric stages couple light into and out of the waveguides. The chip is positioned on the oven for temperature control. 
After coupling the light out of waveguides, a free-space arrangement of three short-pass dichroic mirrors separates the pump (near 1557~nm), signal (near 1078 nm), and sum-frequency output (near 637~nm). InGaAs and Si photodetectors are used to measure the power at the pump, input signal, and converted signal wavelengths. A broadband SNSPD from ID Quantique covering both visible and near-IR wavelengths is used for noise measurement.

\textbf{Calibration of loss, power, and noise.}
We independently measure the device transmission using waveguides of different lengths at 637~nm, 1078~nm, and 1557~nm. 
Linear fit of the transmission in decibel with respect to the waveguide length is performed. The propagation loss and fiber-waveguide coupling efficiency can be extracted from the fitted slope and intercept. 
The measured propagation losses $\alpha$ are $2.4\pm0.5$~dB/cm, $0.6\pm0.4$~dB/cm, and $0.5\pm0.1$~dB/cm, respectively.
The measured fiber-waveguide coupling efficiencies $T_{\mathrm{coupling}}$ are $27\pm2\%$, $33\pm2\%$, $36\pm1\%$ at 637~nm, 1078~nm, and 1557~nm respectively.
The transmissions of free-space WDM ($T_{\mathrm{WDM}}$) are $50\pm4\%$, $54\pm5\%$, and $59\pm3\%$ at 637~nm, 1078~nm, and 1557~nm respectively.
The SNSPD has a nominal efficiency of 77\% and 80\% efficiencies at 637~nm and 1078~nm respectively.

On-chip input power for the pump and input signal is estimated with $P_\mathrm{in}\cdot T_{\mathrm{coupling}}$ with $P_\mathrm{in}$ the power in the input fiber. On-chip output SFG power is estimated based on $\frac{P_\mathrm{PD}}{T_{\mathrm{WDM}}T_{\mathrm{coupling}}}$ with $P_\mathrm{PD}$ the power measured by the photodetector. On-chip noise rate is calculated as $\frac{N}{T_{\mathrm{coupling}} \cdot T_{\mathrm{WDM}} \cdot T_{\mathrm{SNSPD}}}$ with $N$ the photon count rate on the SNSPD.

\bibliography{bibliography}

@article{castelletto2020silicon,
  title={Silicon carbide color centers for quantum applications},
  author={Castelletto, Stefania and Boretti, Alberto},
  journal={Journal of Physics: Photonics},
  volume={2},
  number={2},
  pages={022001},
  year={2020},
  publisher={IOP Publishing}
}

@article{janitz2020cavity,
  title={Cavity quantum electrodynamics with color centers in diamond},
  author={Janitz, Erika and Bhaskar, Mihir K and Childress, Lilian},
  journal={Optica},
  volume={7},
  number={10},
  pages={1232--1252},
  year={2020},
  publisher={Optical Society of America}
}

@article{kumar1990quantum,
  title={Quantum frequency conversion},
  author={Kumar, Prem},
  journal={Optics letters},
  volume={15},
  number={24},
  pages={1476--1478},
  year={1990},
  publisher={Optical Society of America}
}

@article{doi:10.1073/pnas.1419326112,
  author = {Kurizki, Gershon and Bertet, Patrice and Kubo, Yuimaru and M{\o}lmer, Klaus and Petrosyan, David and Rabl, Peter and Schmiedmayer, J{\"{o}}rg},
  title = {{Quantum technologies with hybrid systems}},
  journal = {Proceedings of the National Academy of Sciences},
  year = {2015},
  volume = {112},
  number = {13},
  pages = {3866--3873},
  doi = {10.1073/pnas.1419326112},
  url = {https://doi.org/10.1073/pnas.1419326112},
}

@article{Abobeih_2018,
  author = {Abobeih, M. H. and Cramer, J. and Bakker, M. A. and Kalb, N. and Markham, M. and Twitchen, D. J. and Taminiau, T. H.},
  title = {{One-second coherence for a single electron spin coupled to a multi-qubit nuclear-spin environment}},
  journal = {Nature Communications},
  year = {2018},
  volume = {9},
  number = {1},
  pages = {2552},
  doi = {10.1038/s41467-018-04916-z},
  url = {https://doi.org/10.1038/s41467-018-04916-z},
}

@article{MA201269,
  author = {Ma, Lijun and Slattery, Oliver and Tang, Xiao},
  title = {{Single photon frequency up-conversion and its applications}},
  journal = {Physics Reports},
  year = {2012},
  volume = {521},
  number = {2},
  pages = {69--94},
  doi = {10.1016/j.physrep.2012.07.006},
  url = {https://doi.org/10.1016/j.physrep.2012.07.006},
}

@article{Pelc:11,
  author = {Pelc, J. S. and Ma, L. and Phillips, C. R. and Zhang, Q. and Langrock, C. and Slattery, O. and Tang, X. and Fejer, M. M.},
  title = {{Long-wavelength-pumped upconversion single-photon detector at 1550 nm: performance and noise analysis}},
  journal = {Optics Express},
  year = {2011},
  volume = {19},
  number = {22},
  pages = {21445},
  doi = {10.1364/OE.19.021445},
  url = {https://doi.org/10.1364/OE.19.021445},
}

@article{PhysRevApplied.14.034035,
  author = {Zheng, Ming-Yang and Yao, Quan and Wang, Bin and Xie, Xiu-Ping and Zhang, Qiang and Pan, Jian-Wei},
  title = {{Integrated Multichannel Lithium Niobate Waveguides for Quantum Frequency Conversion}},
  journal = {Physical Review Applied},
  year = {2020},
  volume = {14},
  number = {3},
  pages = {034035},
  doi = {10.1103/PhysRevApplied.14.034035},
  url = {https://doi.org/10.1103/PhysRevApplied.14.034035},
}

@article{Yu:20,
  author = {Yu, Yong and Ma, Fei and Luo, Xi-Yu and Jing, Bo and Sun, Peng-Fei and Fang, Ren-Zhou and Yang, Chao-Wei and Liu, Hui and Zheng, Ming-Yang and Xie, Xiu-Ping and Zhang, Wei-Jun and You, Li-Xing and Wang, Zhen and Chen, Teng-Yun and Zhang, Qiang and Bao, Xiao-Hui and Pan, Jian-Wei},
  title = {{Entanglement of two quantum memories via fibres over dozens of kilometres}},
  journal = {Nature},
  year = {2020},
  volume = {578},
  number = {7794},
  pages = {240--245},
  doi = {10.1038/s41586-020-1976-7},
  url = {https://doi.org/10.1038/s41586-020-1976-7},
}

@article{Xu:07,
  author = {Xu, Hai and Ma, Lijun and Mink, Alan and Hershman, Barry and Tang, Xiao},
  title = {{1310-nm quantum key distribution system with up-conversion pump wavelength at 1550 nm}},
  journal = {Optics Express},
  year = {2007},
  volume = {15},
  number = {12},
  pages = {7247},
  doi = {10.1364/OE.15.007247},
  url = {https://doi.org/10.1364/OE.15.007247},
}

@article{Rakher:10,
  author = {Rakher, Matthew T. and Ma, Lijun and Slattery, Oliver and Tang, Xiao and Srinivasan, Kartik},
  title = {{Quantum transduction of telecommunications-band single photons from a quantum dot by frequency upconversion}},
  journal = {Nature Photonics},
  year = {2010},
  volume = {4},
  number = {11},
  pages = {786--791},
  doi = {10.1038/nphoton.2010.221},
  url = {https://doi.org/10.1038/nphoton.2010.221},
}

@article{Kuo:18,
  author = {Kuo, Paulina S. and Pelc, Jason S. and Langrock, Carsten and Fejer, M. M.},
  title = {{Using temperature to reduce noise in quantum frequency conversion}},
  journal = {Optics Letters},
  year = {2018},
  volume = {43},
  number = {9},
  pages = {2034},
  doi = {10.1364/OL.43.002034},
  url = {https://doi.org/10.1364/OL.43.002034},
}

@article{PhysRevLett.124.010510,
  author = {van Leent, Tim and Bock, Matthias and Garthoff, Robert and Redeker, Kai and Zhang, Wei and Bauer, Tobias and Rosenfeld, Wenjamin and Becher, Christoph and Weinfurter, Harald},
  title = {{Long-Distance Distribution of Atom-Photon Entanglement at Telecom Wavelength}},
  journal = {Physical Review Letters},
  year = {2020},
  volume = {124},
  number = {1},
  pages = {010510},
  doi = {10.1103/PhysRevLett.124.010510},
  url = {https://doi.org/10.1103/PhysRevLett.124.010510},
}

@article{photonics9120971,
  author = {Guo, Miao and Zhang, Kong and Zhang, Yunhao and He, Jun and Wang, Junmin},
  title = {{Improving the Signal-to-Noise Ratio of Photonic Frequency Conversion from 852 nm to 1560 nm Based on a Long-Wavelength Laser-Pumped PPLN Waveguide Module}},
  journal = {Photonics},
  year = {2022},
  volume = {9},
  number = {12},
  pages = {971},
  doi = {10.3390/photonics9120971},
  url = {https://doi.org/10.3390/photonics9120971},
}

@article{Maring:2018,
  author = {Maring, Nicolas and Lago-Rivera, Dario and Lenhard, Andreas and Heinze, Georg and de Riedmatten, Hugues},
  title = {{Quantum frequency conversion of memory-compatible single photons from 606  nm to the telecom C-band}},
  journal = {Optica},
  year = {2018},
  volume = {5},
  number = {5},
  pages = {507},
  doi = {10.1364/optica.5.000507},
  url = {https://doi.org/10.1364/optica.5.000507},
}

@article{Gao:2015,
  author = {Gao, W. B. and Imamoglu, A. and Bernien, H. and Hanson, R.},
  title = {Coherent manipulation, measurement and entanglement of individual solid-state spins using optical fields},
  journal = {Nature Photonics},
  volume = {9},
  pages = {363--373},
  year = {2015},
  doi = {10.1038/nphoton.2015.58},
  url = {https://www.nature.com/articles/nphoton.2015.58}
}

@article{Blatt:08,
  author = {Blatt, Rainer and Wineland, David},
  title = {Entangled states of trapped atomic ions},
  journal = {Nature},
  volume = {453},
  pages = {1008--1015},
  year = {2008},
  doi = {10.1038/nature07125},
  url = {https://www.nature.com/articles/nature07125}
}

@article{Specht:11,
  author = {Specht, Holger P. and N{\"o}lleke, Christian and Reiserer, Andreas and Uphoff, Manuel and Figueroa, Eden and Ritter, Stephan and Rempe, Gerhard},
  title = {A single-atom quantum memory},
  journal = {Nature},
  volume = {473},
  pages = {190--193},
  year = {2011},
  doi = {10.1038/nature09997},
  url = {https://www.nature.com/articles/nature09997}
}

@article{Zhu:21,
  author = {Zhu, Di and Shao, Linbo and Yu, Mengjie and Cheng, Rebecca and Desiatov, Boris and Xin, C. J. and Hu, Yaowen and Holzgrafe, Jeffrey and Ghosh, Soumya and Shams-Ansari, Amirhassan and Puma, Eric and Sinclair, Neil and Reimer, Christian and Zhang, Mian and Lon{\v{c}}ar, Marko},
  title = {{Integrated photonics on thin-film lithium niobate}},
  journal = {Advances in Optics and Photonics},
  year = {2021},
  volume = {13},
  number = {2},
  pages = {242},
  doi = {10.1364/AOP.411024},
  url = {https://doi.org/10.1364/AOP.411024},
}

@article{Wallquist:2009,
  author = {Wallquist, M and Hammerer, K and Rabl, P and Lukin, M and Zoller, P},
  title = {{Hybrid quantum devices and quantum engineering}},
  journal = {Physica Scripta},
  year = {2009},
  volume = {T137},
  pages = {014001},
  doi = {10.1088/0031-8949/2009/T137/014001},
  url = {https://doi.org/10.1088/0031-8949/2009/T137/014001},
}

@article{Wang:2023,
  author = {Wang, Xina and Jiao, Xufeng and Wang, Bin and Liu, Yang and Xie, Xiu-Ping and Zheng, Ming-Yang and Zhang, Qiang and Pan, Jian-Wei},
  title = {{Quantum frequency conversion and single-photon detection with lithium niobate nanophotonic chips}},
  journal = {npj Quantum Information},
  year = {2023},
  volume = {9},
  number = {1},
  pages = {38},
  doi = {10.1038/s41534-023-00704-w},
  url = {https://doi.org/10.1038/s41534-023-00704-w},
}

@article{PhysRevApplied.9.064031,
  author = {Dr{\'{e}}au, Ana{\"{i}}s and Tcheborateva, Anna and Mahdaoui, Aboubakr El and Bonato, Cristian and Hanson, Ronald},
  title = {{Quantum Frequency Conversion of Single Photons from a Nitrogen-Vacancy Center in Diamond to Telecommunication Wavelengths}},
  journal = {Physical Review Applied},
  year = {2018},
  volume = {9},
  number = {6},
  pages = {064031},
  doi = {10.1103/PhysRevApplied.9.064031},
  url = {https://doi.org/10.1103/PhysRevApplied.9.064031},
}

@article{adapted,
  author = {Chen, Pao-Kang and Briggs, Ian and Cui, Chaohan and Zhang, Liang and Shah, Manav and Fan, Linran},
  title = {{Adapted poling to break the nonlinear efficiency limit in nanophotonic lithium niobate waveguides}},
  journal = {Nature Nanotechnology},
  year = {2024},
  volume = {19},
  number = {1},
  pages = {44--50},
  doi = {10.1038/s41565-023-01525-w},
  url = {https://doi.org/10.1038/s41565-023-01525-w},
}

@article{Fan:21,
  author = {Fan, Heng and Ma, Zhaohui and Chen, Jiayang and Li, Zhan and Tang, Chao and Sua, Yong Meng and Huang, Yuping},
  title = {{Photon conversion in thin-film lithium niobate nanowaveguides: a noise analysis}},
  journal = {Journal of the Optical Society of America B},
  year = {2021},
  volume = {38},
  number = {7},
  pages = {2172},
  doi = {10.1364/JOSAB.425318},
  url = {https://doi.org/10.1364/JOSAB.425318},
}

@misc{vonChamier:25,
  author = {von Chamier, Fabrice and Hanel, Joscha and M{\"u}ller, Chris and Li, Wanrong and K{\"o}gler, Roger Alfredo and Benson, Oliver},
  title = {{Low-Noise Cascaded Frequency Conversion of 637.2 nm Light to the Telecommunication C-Band in a Single-Waveguide Device}},
  year = {2025},
  eprint = {2502.14557},
  archivePrefix = {arXiv},
  primaryClass = {quant-ph},
  note = {Version 1; numerical comparison uses the author preprint},
  url = {https://arxiv.org/abs/2502.14557v1}
}

@misc{Kao:26,
  author = {Kao, Yung-Cheng and Huang, Jiaqi and Briggs, Ian and Chen, Pao-Kang and Fan, Linran},
  title = {{Efficient net-gain integrated optical parametric amplifier in the quantum regime}},
  year = {2026},
  eprint = {2602.05982},
  archivePrefix = {arXiv},
  primaryClass = {physics.optics},
  note = {Preprint, version 1},
  doi = {10.48550/arXiv.2602.05982},
  url = {https://arxiv.org/abs/2602.05982v1}
}
\end{document}